\documentclass[preprints,article,submit,moreauthors]{Definitions/mdpi}

\usepackage{siunitx}
\firstpage{1} 
\pubvolume{1}
\issuenum{1}
\articlenumber{0}
\pubyear{2022}
\copyrightyear{2022}
\datereceived{} 
\dateaccepted{} 
\datepublished{} 
\hreflink{https://doi.org/} 
\pdfoutput=1

\Title{Dynamic image recognition in a spiking neuron network supplied by astrocytes}

\TitleCitation{Dynamic image recognition in a spiking neuron network supplied by astrocytes}

\Author{Sergey V. Stasenko $^{1}$*\orcidA{},  and Victor B. Kazantsev $^{1}$\orcidA{}}

\AuthorNames{Sergey V. Stasenko and Victor B. Kazantsev}

\AuthorCitation{Stasenko, S.V.; Kazantsev V.B.}

\address{%
$^{1}$ \quad Moscow Institute of Physics and Technology\\
}

\corres{Correspondence: stasenko@neuro.nnov.ru}

\abstract{Mathematical model of spiking neuron network (SNN) supplied by astrocytes is investigated. The astrocytes are specific type of brain cells which are not electrically excitable but inducing chemical modulations of neuronal firing. We analyzed how the astrocytes influence on images encoded in the form of dynamic spiking pattern of the SNN. Serving at much slower time scale the astrocytic network interacting with the spiking neurons can remarkably enhance the image recognition quality. Spiking dynamics was affected by noise distorting the information image. We demonstrated that the activation of astrocyte can significantly suppress noise influence improving dynamic image representation by the SNN.}


\begin{document}
\section{Introduction}

The construction of biologically relevant models of brain information processing  still remains one of the key tasks of modern mathematical neuroscience. In neurobiology, key mechanisms of information processing concern synaptic transmission between the brain network neurons. Synaptic plasticity, e.g. adaptive changes in the connection strengths, is believed to be the main instrument of implementation learning and memory in the neuron networks. Following the neurobiological studies many mathematical models targeted to describe experimental results and, hence, to imitate brain functions have been proposed. However, it is still remain a challenge on how at network level brain circuits can generate so finely tuned and effective information representation and processing. 

In recent two decades neurobiological experiments have revealed that neurons and neuronal networks are not alone in the brain universe.
It was found that glial cells, particularly astrocytes, known before as just ``supporting'' cells providing mostly metabolic functions, can also participate in information processing by means of chemical regulations of neuronal activity and synaptic transmission \cite{Araque1998, Araque1999, Wittenberg2002, Wang1999}. Inclusion of the third player, e.g. astrocytes, in the classical “presynapse-postsynapse” signal transmission scheme led to the concept of a tripartite synapse \cite{Wittenberg2002, Araque1999, Haydon2001}. Astrocytes through calcium-dependent release of neuroactive chemicals (for example, glutamate) affect the pre- and postsynaptic compartments of the synapse. When spikes are generated by a presynaptic neuron, a neurotransmitter (for example, glutamate) is released from the presynaptic terminal. By diffusion part of the chemicals leave synaptic cleft and bind to metabotropic glutamate receptors (mGluRs) on the astrocyte, which may be located near the presynaptic terminal. Activation of metabotropic glutamate receptors G-mediated leads to the formation of inositol-1,4,5-triphosphate (IP$_3$). This process after a cascade of molecular transformations inside the astrocyte leads to the release of $Ca^{2+}$ into the cytoplasm. It induces the release of the neuroactive chemicals called gliatransmitters (for example, glutamate, adenosine triphosphate (ATP), D-serine, GABA) back to the extrasynaptic space. Next, they bind to pre- or postsynaptic receptors resulting finally in modulation of the efficiency of synaptic transmission completing the feedback loop \cite{Perea2009}.

Many mathematical models were then proposed to explore the functional role of astrocytes in neuronal dynamics. They include model of the “dressed neuron,” which describes the astrocyte- mediated changes in neural excitability \cite{Nadkarni2004,Nadkarni2007}, model of the astrocyte serving as a frequency selective “gate keeper” \cite{Volman2007}, model of the astrocyte regulating presynaptic functions \cite{DePitta2011} and many others. In particular, it was demonstrated that gliotransmitters can effectively con trol presynaptic facilitation and depression. The model of the tripartite synapse has recently been employed to demonstrate the functions of astrocytes in the coordination of neuronal network signaling, in particular, spike-timing-dependent plasticity and learning \cite{Postnov2007, Amiri2011, Wade2011}. In models of astrocytic networks, communication between astrocytes has been described as $Ca^{2+}$ wave propagation and synchronization of $Ca^{2+}$ waves \cite{Ullah2006, Kazantsev2009}. However, due to a variety of potential actions, that may be specific for brain regions and neuronal sub-types, the functional roles of astrocytes in network dynamics are still a subject of debate.

Role of astrocytes as collaborators of spiking neuron networks (SNN) in implementing learning and memory functions have been intensivly discussed in recent computational models \cite{gordleeva2021modeling, tsybina2022astrocytes,gordleeva2022situation}. Specifically, it was demonstrated that the astrocytes serving at much slower time scale can help SNN to distinguish highly overlapping images. Here we present another SNN model accompanied by the astrocytes that 
can significantly enhance recognition of information images encoded in the form of dynamical spiking patterns stored by the SNN.

\section{The model}

\subsection{Mathematical model of single neuron}
The SNN's individual neuron is described by the Hodgkin-Huxley model \cite{Hodgkin1952,Izhikevich2007} determined that the squid axon curries three major currents: voltage-gated persistent $K^+$ current, $I_K$, with four activation gates (resulting in the term $n^4$ in the equation below, where $n$ is the activation variable for $K^+$), voltage-gated transient $Na^+$ current, $I_{Na}$, with three activation gates and one inactivation gate (term $m^3 h$ below), and Ohmic leak current, $I_L$, which is carried mostly by $Cl^ -$ ions. The complete set (Eq. \eqref{eq:hh}) of space-clamped Hodgkin-Huxley equations is
\begin{eqnarray}
C\dot{V} &=& I_{inj} - \overbrace{\bar{g}_{Na}m^3h(V-V_{Na})}^{I_{Na}}- 
\overbrace{\bar{g}_Kn^4(V-V_K)}^{I_K} - 
\overbrace{\bar{g}_L (V-V_L)}^{I_L} \notag\\
\dot{n} &=& \alpha_n(V) (1-n) - \beta_n(V)n \notag\\
\dot{m} &=& \alpha_m(V) (1-m) - \beta_m(V)m \notag\\
\dot{h} &=& \alpha_h(V) (1-h) - \beta_h(V)h, \label{eq:hh}
\end{eqnarray}
  where:
\begin{eqnarray}
I_{inj} &=& I_{stim} + I_{noise} + I_{syn} \label{eq:I}\\
\alpha_n(V) &=& \frac{0.01(V+55)}{1-exp[-(V+55)/10]} \notag\\
\beta_n(V) &=& 1.125exp[-(V+65)/80]\notag\\
\alpha_m(V) &=& \frac{0.1(V+40)}{1-exp[-(V+40)/10]} \notag\\
\beta_m(V) &=& 4exp[-(V+65)/18]\notag\\
\alpha_h(V) &=& 0.07exp[-(V+65)/20] \notag\\
\beta_n(V) &=& \frac{1}{1+exp[-(V+35)/10]}\notag
\end{eqnarray}

Shifted Nernst equilibrium potentials for $I_{Na}$, $I_K$ and $I_L$ are 
$V_{Na} = \SI{50}{\mV}$, $V_{K} = \SI{-77}{\mV}$ and $V_{L} = \SI{-54.4}{\mV}$, respectively. Typical values of maximal conductances for $I_{Na}$, $I_K$ and $I_L$ are $\bar{g}_{Na} = \SI{36}{mS/cm^2}$, $\bar{g}_K = \SI{120}{mS/cm^2}$ and $\bar{g}_L = \SI{0.3}{mS/cm^2}$, respectively. The functions $\alpha(V)$ and $\beta(V)$ describe the transition rates between open and closed states of the channels. $C = \SI{1}{\mu F/cm^2}$ is the membrane capacitance and $I_{inj}$ (Eq.\eqref{eq:I}) is the applied current which consists from three parts:  $I_{stim}$, $I_{noise}$ and $I_{syn}$. 

\subsection{Applied currents}

Images applied to the SNN were encoded as matrices $M$ of size $n \times  k$ and values from 0 to 1 for each pixel, where 0 is the absence of color, and $n$ and $k$ are the corresponding image sizes (length and width). Next, the matrix $M$ was transformed into an $l \times 1 $  vector $S$, where $l = n \times k$ and corresponds to the neuron index in the neural network. Thus, the stimulation current, $I_{stim}$, will be written in the following form:
\begin{eqnarray}
I_{stim} &=& S \times A_S, \label{eq:Istim}
\end{eqnarray}
where $A_S$ is the amplitude of stimulus taken here for illustration with value $\SI{5.3}{nA}$.

The synaptic current,$I_{syn}$, is modeled using conductance-based approach as following form:
\begin{eqnarray}
I_{syn} &=& g_{j}(V_{j} - V), \label{eq:Isyn}
\end{eqnarray}
where:
\begin{eqnarray}\label{eq:gdot}
\dot{g_{j}} &=& \frac{-g_{j}}{\tau_{j}} 
\end{eqnarray}
In our model index $j$ is used for excitatory (exc) and inhibitory (inh) synapses. Reversal potentials for synaptic currents are equal $\SI{0}{mV}$ and $\SI{-80}{mV}$ for excitatory and inhibitory synapses, respectively. $\tau_{j}$ is the time relaxation equaled $\SI{5}{ms}$ and $\SI{10}{ms}$ for excitatory and inhibitory synapses, respectively.
Excitatory (inhibitory) synapses will increase the excitatory (inhibitory) conductance in the postsynaptic cell whenever a presynaptic action potential arrives:
\begin{eqnarray}\label{eq:g}
g_{j} \leftarrow g_{j} + w_{j},
\end{eqnarray}
where $w_{j}$ - synaptic weight equaled $\SI{3}{nS}$ and $\SI{77}{nS}$ for excitatory and inhibitory synapses, respectively.

Besides the synaptic input, each neuron receives a noisy thalamic input ($I_{noise}$). The noisy thalamic input is set in a random way for all neurons in the range from 0 to $A_{noise}$.

Each spike in the neuron model induces the release of neurotransmitter. To describe the neuron to astrocyte cross-talk, here we only focus on the excitatory neurons releasing glutamate. Following earlier experimental and modeling studies, we assumed that the glutamate-mediate exchange was the key mechanism to induce coherent neuronal excitations \cite{Angulo2004, Halassa2009}. The role of GABAergic neurons in our network is to support the excitation and inhibition balance avoiding hyperexcitation states.

For simplicity, we take a phenomenological model of released glutamate dynamics. In the mean field approximation average concentration of synaptic glutamate concentration for each excitatory synapses, $X_{e}$, was described by this equations:
\begin{equation}
X_{e}(t) = 
\left\{
\begin{array}{lcl}
X_{e}(t_s)exp(-t/\tau_X),  & \mbox{if} & t_s<t<t_{s+1},  \\
    X_{e}(t_s-0)+1, & \mbox{if} & t=t_s,
\end{array}
 \right.
  \label{eq:X}
\end{equation}
where $e=1,2,3,.$ is the index of excitatory presynaptic neurons, $s=1,2,3,\ldots$ is the index of the presynaptic spikes, $\tau_{X}$ is the time relaxation equal $\SI{80}{ms}$. After the spike is generated on the presynaptic neuron, the neurotransmitter is released.

\subsection{Tripartite synapses}
To describe the dynamics of a tripartite synapse, we used the mean-field approach to describe changes in the concentration of neuroactive substances (neurotransmitter and gliatransmitter), proposed in the work \cite{Gordleeva2012}. In the proposed model, the filtering of the external noise signal applied to neurons is carried out due to synaptic depression, which consists in a decrease in the probability of neurotransmitter release, which in turn leads to a decrease in the strength of the connection between neurons (work \cite{Lazarevich2017}).
Part of the synaptic glutamate can bind to metabotropic glutamate receptors of the astrocyte processes. Next, after a cascade of molecular transformations mediated by elevation of intracellular calcium the astrocyte release of gliatransmitter back to the extracellular space. For our purpose, in mathematical model we dropped detailed description of these transformations, defining only input-output functional relation between the neurotransmitter and gliatransmitter concentrations in the following form \cite{Stasenko2020,Gordleeva2012}: 
\begin{equation}
\dot{dY_{e}} = - Y_{e}/\tau_{Y_{e}} + \frac{\beta_{Y}}{1+exp(-X_{e} + X_{thr})} \label{eq:Y}
\end{equation}
where $e=1,2,3,\ldots$ is the index of excitatory neuron,  $Y_e$ is the gliatransmitter concentration in the neighborhood of corresponding excitatory synapse, $\tau_{Y_{e}}$ is the clearance rate equal $\SI{120}{ms}$. 
The second term in Eq. (\ref{eq:Y}) describes the gliatransmitter production ($\beta_{Y} = 1$) when the mean field concentration of gliatransmitter exceeds some threshold, $X_{thr}$, equal to 1. Figure \ref{fig:scheme} illustrates the network construction and neuron to astrocyte crosstalk for excitatory glutamatergic synapses. 

It follows from experimental facts that astrocytes can influence on probability of neurotransmitter release \cite{Martin2007, Jourdain2007,Fiacco2004}. 
In turn, it results in modulation synaptic currents. We accounted this in the following form for glutamatergic synapses: 
\begin{equation}
w_{ext} \leftarrow w_{ext} (1- \frac{\gamma_{Y}}{(1 + exp(-Y + Y_{thr}))}) \label{eq:w}
\end{equation}
$w_{ext}$ is the weight for glutamatergic synapses between neurons, $\gamma_{Y}$ is the coefficient of astrocyte influence on synaptic connection.

\subsection{Neural Network}
\begin{figure}[H]
    \centering
    \includegraphics[scale=0.18]{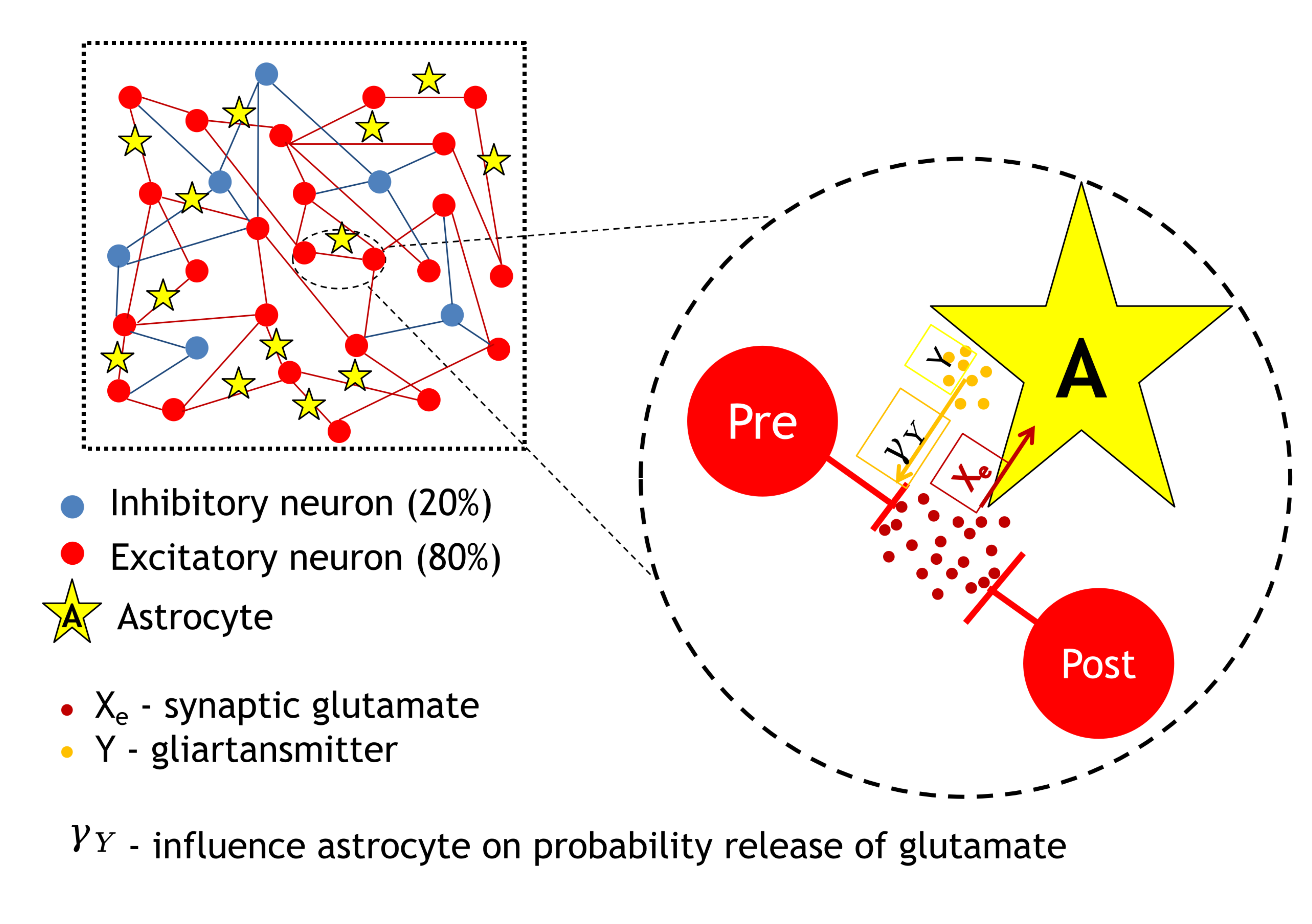}
    \caption{Scheme of neuron-glial network.}
    \label{fig:scheme}
\end{figure}

Schematic representation of network with astrocytic modulation of probability release of neurotransmitter is presented in Fig.\,\ref{fig:scheme}. After the generation of an action potential on the presynaptic neuron, neurotransmitter is released from the presynaptic terminal. Its part can diffuse out of the cleft where it can bind to specific astrocyte receptors \cite{Rusakov1998}. The activation of the astrocyte results in the generation of calcium transients in the form of short-term increase in the intra- cellular concentration of calcium. In turn, the calcium elevations lead to gliotransmitter (particularly glutamate) release. The released gliatransmitter, reaching the presynaptic terminal, leads to a change in the probability of neurotransmitter release, depressing the synaptic current.

The size $N$ of the spike neural network was chosen based on the size of the presented image, i.e. $N = n \times  k$ in real time (resolution 1 ms). Motivated by the anatomy of a mammalian cortex, we choose the ratio of excitatory to inhibitory neurons to be 4 to 1. The probability of connection of excitatory neurons is 5\%, the probability of connection of inhibitory neurons is 10\%. Since the model uses a mean-field approach to describe changes in the main neuroactive substances (neurotransmitter and gliatransmitter), we do not separate the effect of a single astrocyte on a group of neurons or a group of neurons on a single astrocyte, but we introduce into the description of each synaptic contact its own dynamics for the neuro and gliatransmitter.
\section{Results}

\begin{figure}[H]
    \centering
    \includegraphics[scale=0.5]{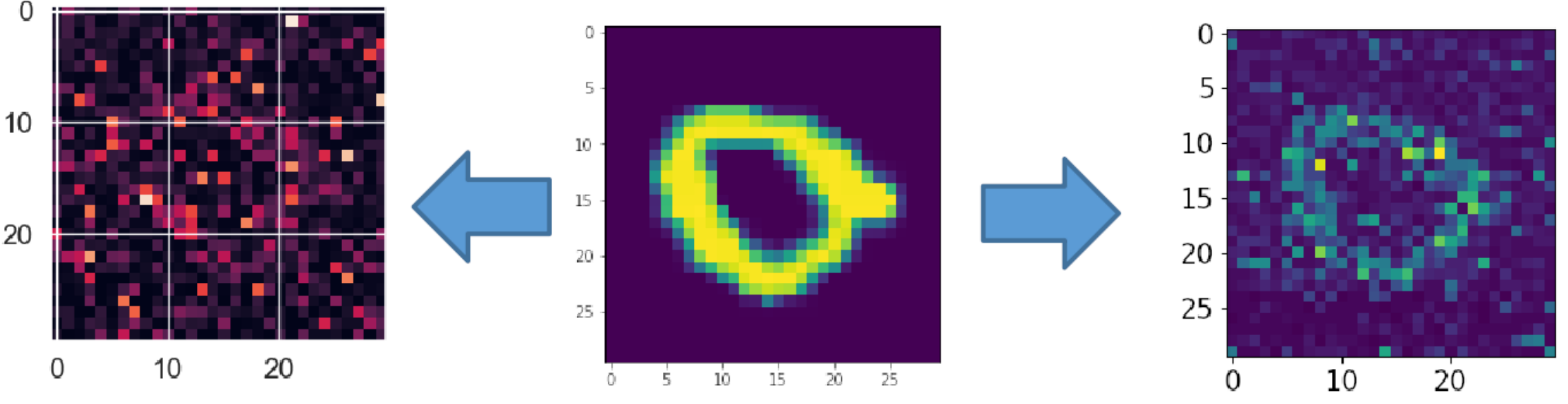}
    \caption{The middle picture is the supplied pattern (zero from handwritten database MNIST) \cite{LeCun2010}, on the left - encoded by the spiking neural network without astrocyte modulation and noise amplitude $A_{noise} = 6$, on the right - with astrocyte modulation and noise amplitude $A_{noise} = 6$.}
    \label{fig:pattern}
\end{figure}

\begin{figure}[H]
    \centering
    \includegraphics[scale=0.3]{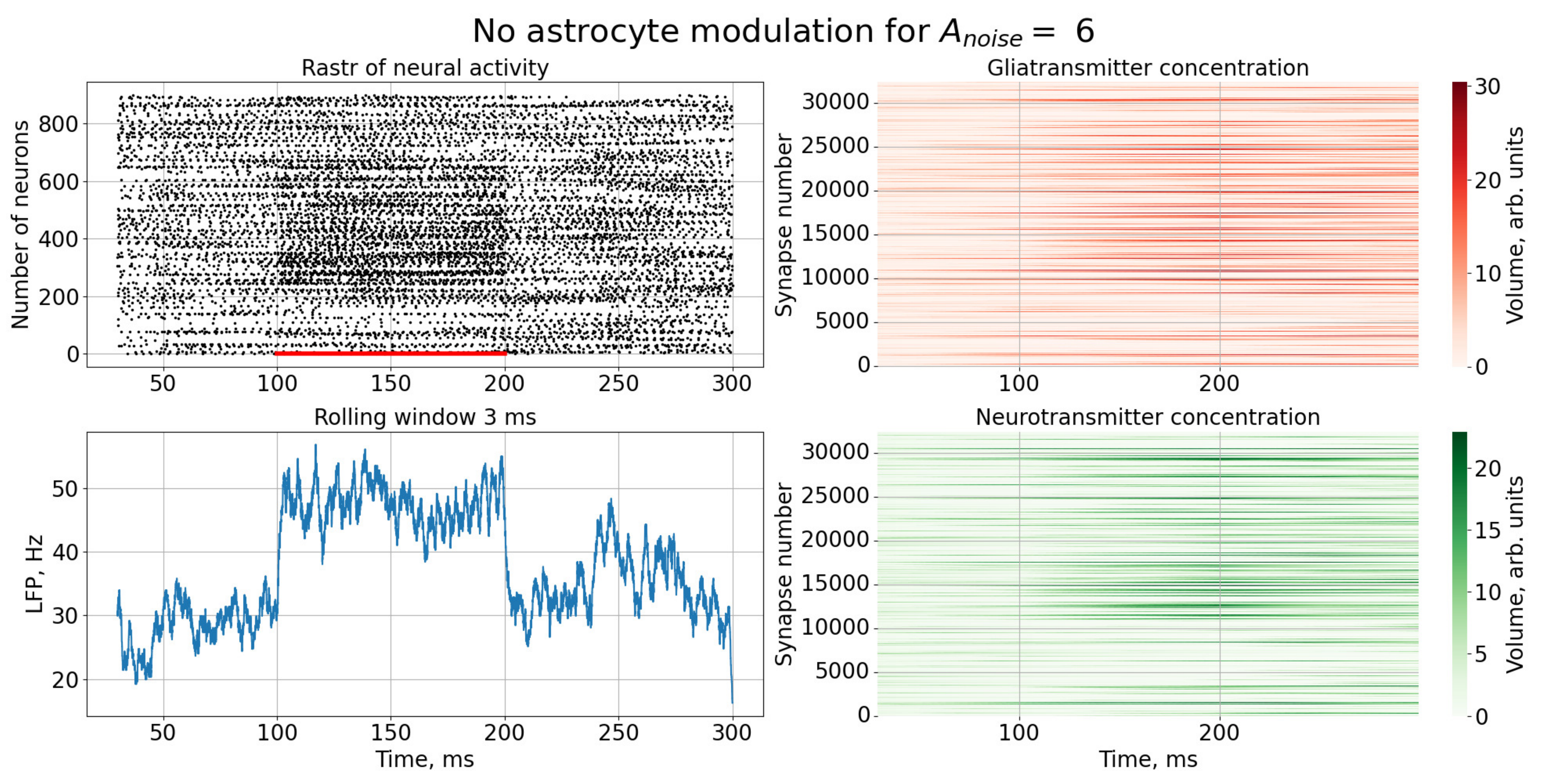}
    \caption{Time series of neural activity (upper left figure in the form of a raster diagram), the corresponding LFP signal with rolling window 3 ms (lower left figure) and time series of concentrations gliatransmitters (upper right figure) and concentration neurotransmitters (lower right figure) in the case of an external noise signal with $A_{noise} = 6$ without astrocytic modulation.}
    \label{fig:series_no_astro}
\end{figure}

\begin{figure}[H]
    \centering
    \includegraphics[scale=0.3]{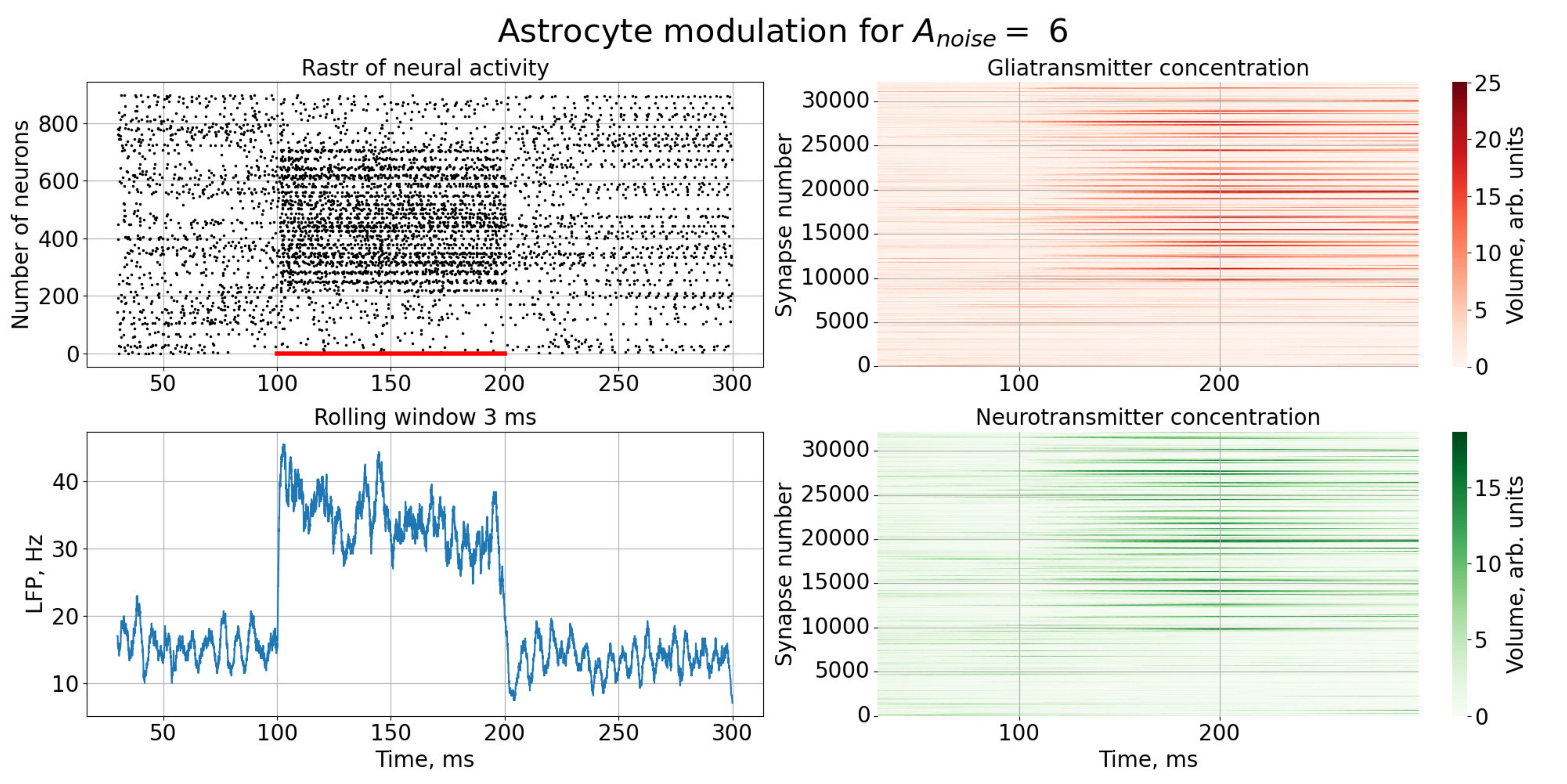}
    \caption{Time series of neural activity (upper left figure in the form of a raster diagram), the corresponding LFP signal with rolling window 3 ms (lower left figure) and time series of concentrations gliatransmitters (upper right figure) and concentration neurotransmitters (lower right figure) in the case of an external noise signal with $A_{noise} = 6$ with astrocytic modulation.}
    \label{fig:series_astro}
\end{figure}

To demonstrate the effect of an astrocyte on the neural activity of a spike neural network, the problem of representing an image by a neural network in the presence of noise was considered. For this purpose, an image in the form of zero (in Fig.\ref{fig:pattern}, middle panel, from the database MNIST \cite{LeCun2010}) was fed to the spiked neural network including tripartite synapses for 100 ms (Fig. \ref{fig:series_no_astro} and Fig. \ref{fig:series_astro}). The image pixels were converted into a current from 0 to a in a spatial sweep to a layer of excitatory and inhibitory neurons.

\begin{figure}[H]
    \centering
    \includegraphics[scale=0.3]{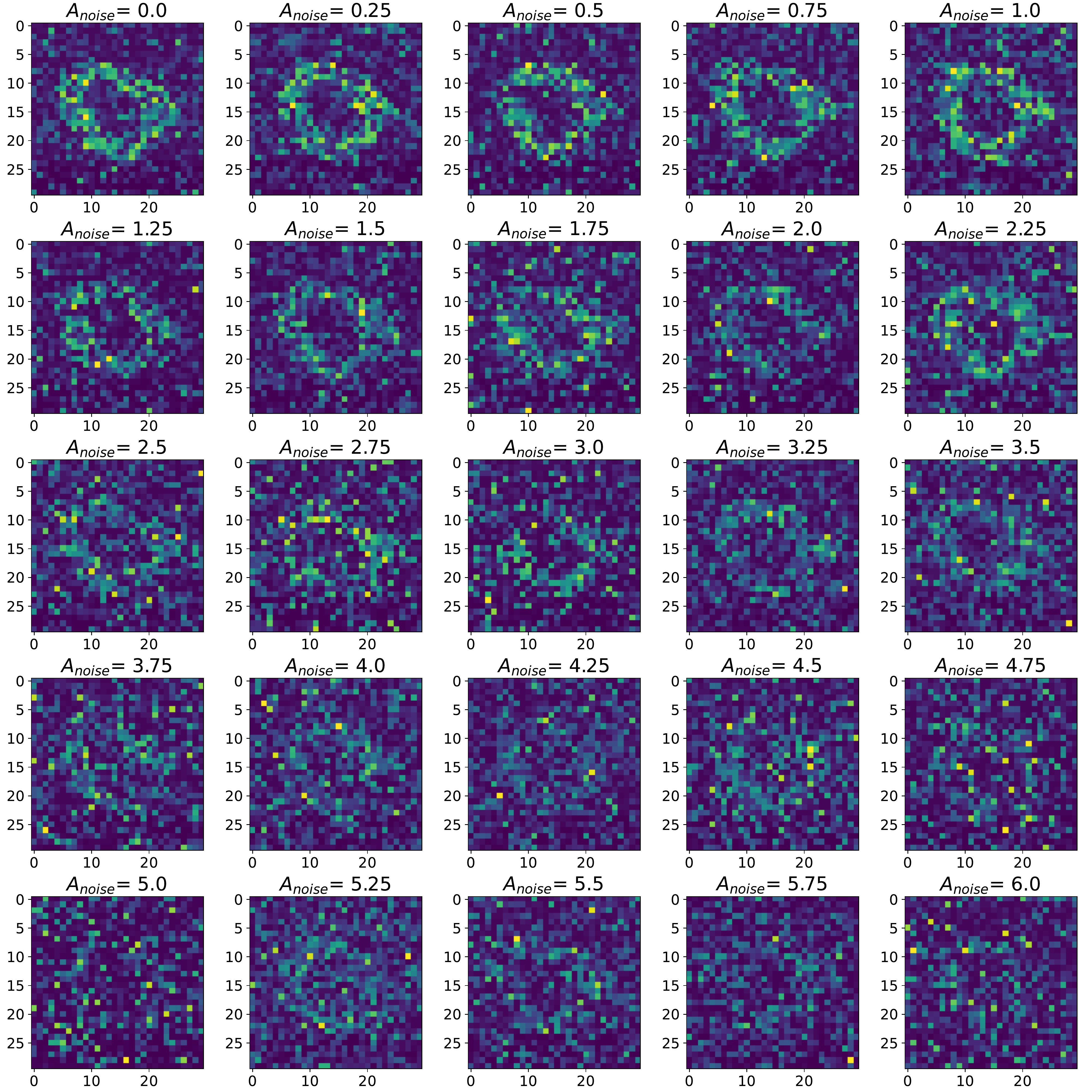}
    \caption{Changes in the spatial sweep of the spike neural network during the representation of the supplied image from the Fig. \ref{fig:pattern} when the amplitude, $A_{noise}$, of the noise current, $I_{noise}$, changes from 0 to 6 without modulation of neuronal activity by astrocytes.}
    \label{fig:rastr_no_astro}
\end{figure}

\begin{figure}[H]
    \centering
    \includegraphics[scale=0.3]{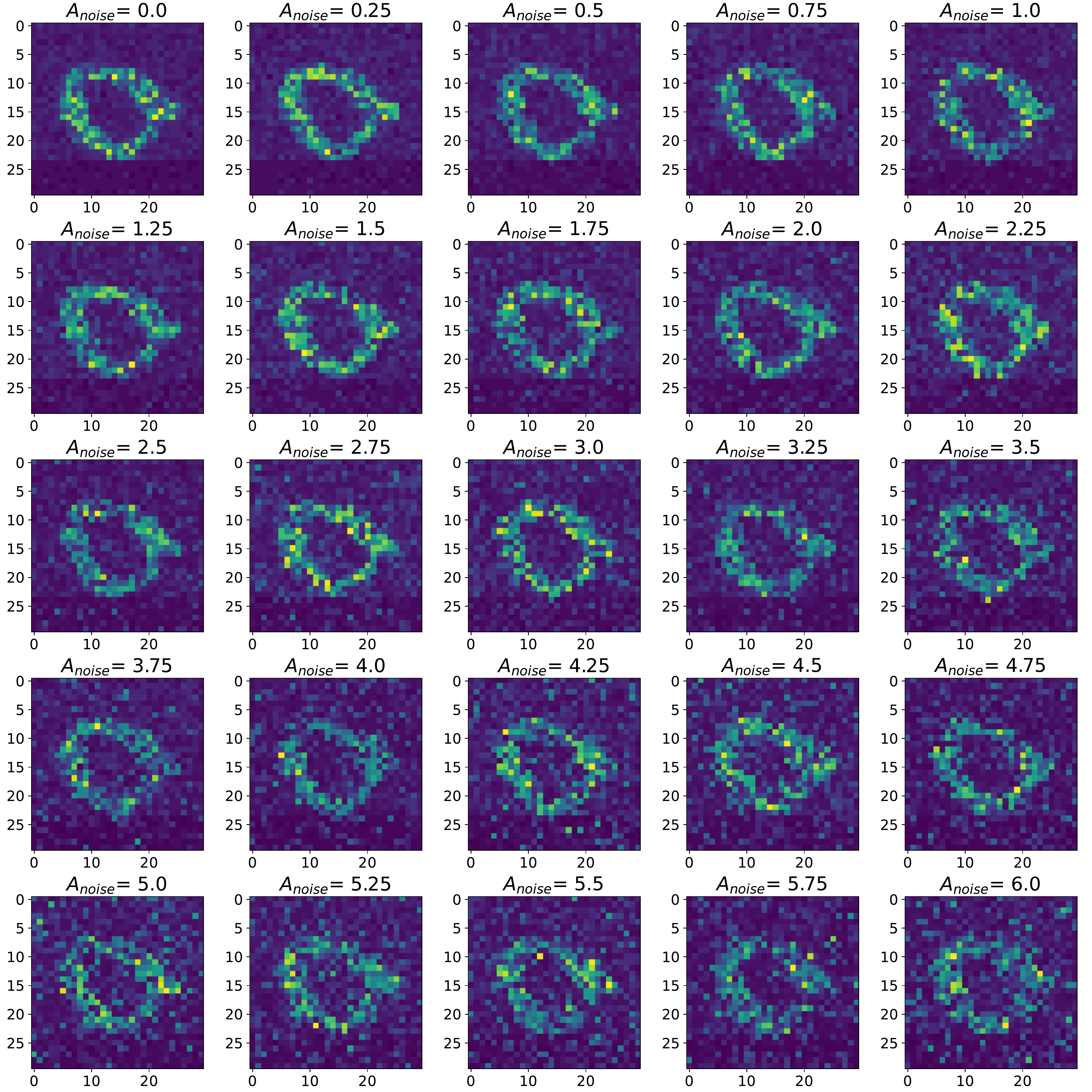}
    \caption{Changes in the spatial sweep of the spike neural network during the representation of the supplied image from the Fig. \ref{fig:pattern} when the amplitude, $A_{noise}$, of the noise current, $I_{noise}$, changes from 0 to 6 with modulation of neuronal activity by astrocytes.}
    \label{fig:rastr_astro}
\end{figure}

A noise signal,$I_{noise}$, is applied to each neuron of the spiking neural network throughout the simulation. As can be replaced, as the amplitude, $A_{noise}$, of the noise signal increases without astrocytic modulation (Fig. \ref{fig:rastr_no_astro}), blurring of the supplied image occurs. Activation of the astrocyte (Fig. \ref{fig:rastr_astro}) through the regulation of neurotransmitter release probability leads to a balancing of excitation and inhibition in the network and thus stabilization of the image representation as the amplitude of the noise signal increases.

\begin{figure}[H]
    \centering
    \includegraphics[scale=0.5]{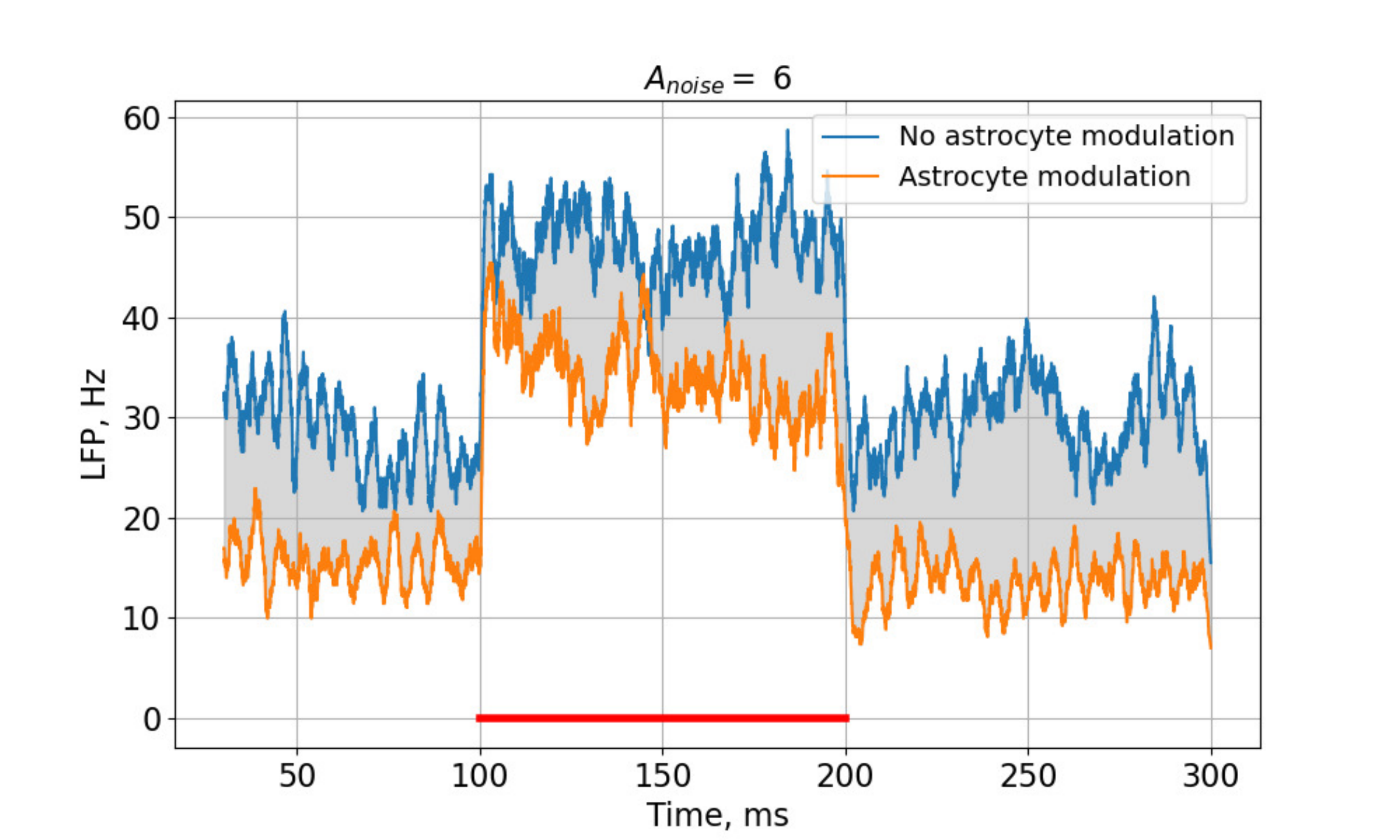}
    \caption{Comparison of the regulation of neuronal activity in the absence and presence of astrocytic modulation for a LFP signal with a noise current amplitude, $A_{noise = 6}$. The red line indicates the time of feeding the image to the neural network.}
    \label{fig:lfp}
\end{figure}

This effect can be most clearly demonstrated by comparing rasters of neural activity (Fig. \ref{fig:series_no_astro} and Fig. \ref{fig:series_astro}) and the corresponding LFP signals (Fig. \ref{fig:lfp}) in the case of a high noise signal amplitude, $A_{noise} = 6$, in the presence and absence of astrocytic modulation. As you can see, the astrocyte lowers the average activity of the neural network (gray area in Fig. \ref{fig:lfp}).

\begin{figure}[H]
    \centering
    \includegraphics[scale=0.4]{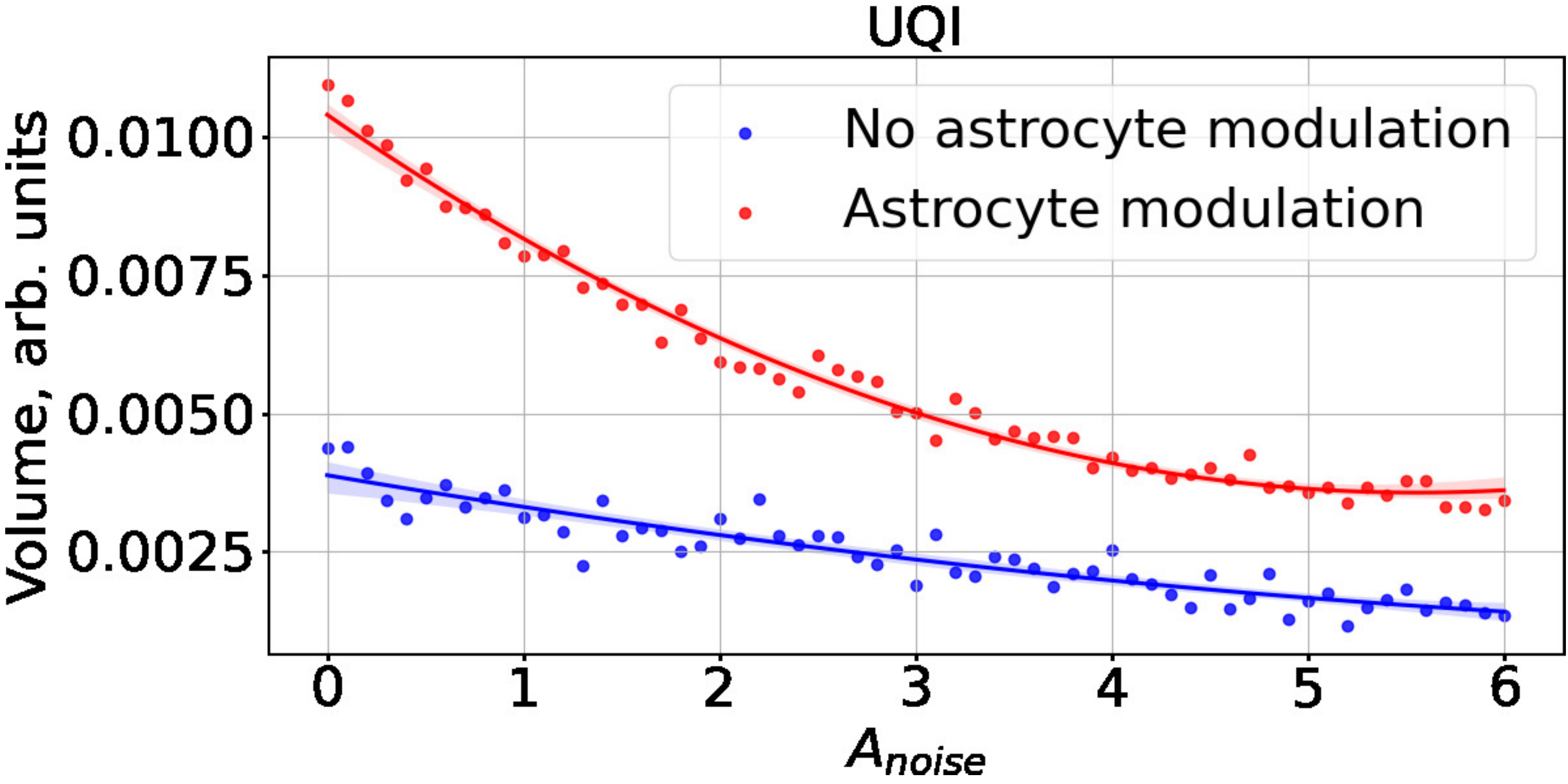}
    \caption{The case of using a quality metric UQI for comparing raster diagrams of neural activity from Fig. \ref{fig:rastr_no_astro} and Fig. \ref{fig:rastr_astro} with an image (middle panel of Fig. \ref{fig:pattern}) fed to a spike neural network with an increase in the amplitude, $A_{noise}$, of the noise signal from 0 to 6 supplied to the neurons of the neural network without astrocytic modulation (blue dots and curve) and with astrocytic modulation (red dots and curve).}
    \label{fig:uqi}
\end{figure}

The figure \ref{fig:uqi} shows a comparison of raster diagrams of neural activity with the image supplied to the neural network over the entire considered range of change in the amplitude, $A_{noise}$, of the noise signal using the quality metric UQI. UQI is image quality technique largely used to evaluate and assess the quality of images and well described at \cite{ZhouWang2002}. This metric is used for modeling any image distortion as a combination of three factors: correlation loss, brightness distortion, and contrast distortion. 
The value of the metric is in the range from 0 to 1, where 1 - the images are completely identical and 0 - completely different. The higher the value of the metric, the more similar the compared images are. As can be seen, in the absence of astrocyte modulation (blue dots and curve in Fig. \ref{fig:uqi}), there is a linear decrease in image similarity. When an astrocyte is activated (red dots and curve in Fig. \ref{fig:uqi}), the most serious drop in similarity is observed in the range of noise signal amplitude values from 0 to 3. Further stabilization occurs.

\section{Discussion}

We constructed SNN model interconnected with astrocytic network. Then, we imposed an arbitrary binary image to the SNN that was kept as {\em dynamic spiking pattern} encoded in the rate of spikes between different neurons. In purely deterministic case it was recognized quite clearly. However, when noise was applied the image was distorted significantly. Activation of astrocytes eventually suppressed the effect of noise preserving the shape of the original image. We calculated changes of image quality characteristics following the UQI metrics demonstrating that the astrocytes were quite effective improving the image quality recognition.

As a point for discussion we feel that astrocytes besides their obvious functional role of low-pass filtering due to the slower time scale of intrinsic process also serve as information processing {\em buffer} capable to store dynamically basic features of the information pattern. Note, that our results are consistent with recent studies of working memory proposed in recent modeling paper https://www.frontiersin.org/articles/10.3389/fncel.2021.631485/full.

\vspace{6pt} 



\authorcontributions{Conceptualization, S.V.S.; methodology, S.V.S.; software, S.V.S.; validation, S.V.S.; formal analysis, S.V.S.; investigation, S.V.S.; resources, S.V.S.; data curation, S.V.S.; writing—original draft preparation, S.V.S. and V.B.K.; writing—review and editing, S.V.S. and V.B.K.; visualization, S.V.S.; supervision, S.V.S.; project administration, S.V.S. and V.B.K.; funding acquisition, S.V.S. and V.B.K. All authors have read and agreed to the published version of the manuscript.
.}

\funding{This research was funded by MIPT Priority 2030 Program.}

\institutionalreview{Not applicable.}

\informedconsent{Not applicable.}

\dataavailability{The data that support the findings of this study are available from the corresponding author upon reasonable request.} 


\conflictsofinterest{The authors declare no conflict of interest.} 




\begin{adjustwidth}{-\extralength}{0cm}

\reftitle{References}

\end{adjustwidth}
\end{document}